\documentclass[aps,prc,twocolumn,showpacs,nofootinbib,superscriptaddress,amsmath,amssymb,showkeys]{revtex4-1}
\usepackage{graphicx}
\usepackage{dcolumn}
\usepackage{bm}
\begin{document}
\title{Combined analysis of the $K^{+}K^{-}$ interaction using near threshold $pp \to ppK^+K^-$ data}
\author{M.~Silarski}
\email[Electronic address: ]{Michal.Silarski@uj.edu.pl}
\affiliation{Institute of Physics, Jagiellonian University, PL-30-059 Cracow, Poland}
\author{P.~Moskal}
\affiliation{Institute of Physics, Jagiellonian University, PL-30-059 Cracow, Poland}
\affiliation{Nuclear Physics Institute, Research Center J{\"u}lich, D-52425 J{\"u}lich, Germany}
\date{\today}
\begin{abstract}
The $K^{+}K^{-}$ final state interaction was investigated based on both the $K^{+}K^{-}$
invariant mass distributions measured at excess
energies of Q~=~10 and~28~MeV and the near threshold excitation function
for the $pp \to ppK^{+}K^{-}$ reaction. The $K^+K^-$ final state enhancement factor was parametrized
using the effective range expansion. 
The effective range of the $K^+K^-$ interaction was estimated to be:
$\mathrm{Re}(b_{K^{+}K^{-}}) = -0.1 \pm 0.4_{stat} \pm 0.3_{sys}~\mathrm{fm}$ and
$\mathrm{Im}(b_{K^{+}K^{-}}) = 1.2^{~+0.1_{stat}~+0.2_{sys}}_{~-0.2_{stat}~-0.0_{sys}}~\mathrm{fm}$,
and the determined real and imaginary parts of the $K^+K^-$ scattering length  amount to:
$\left|\mathrm{Re}(a_{K^{+}K^{-}})\right| = 8.0^{~+6.0_{stat}}_{~-4.0_{stat}}~\mathrm{fm}$
and $\mathrm{Im}(a_{K^{+}K^{-}}) = 0.0^{~+20.0_{stat}}_{~-5.0_{stat}}~\mathrm{fm}$.
\end{abstract}
\maketitle
\section{Introduction}
\label{intro}
The strength of the $K^+K^-$ interaction is a crucial quantity regarding the formation of
a hypothetical kaon--antikaon bound state. Existence of such a state 
could explain the nature of the
$a_0(980)$ and $f_0(980)$ scalar mesons~\cite{Lohse,Weinstein},
whose masses are very close to the sum of the $K^{+}$ and $K^{-}$ masses.\footnote{
Besides the standard
interpretation as $q\bar{q}$ mesons~\cite{Morgan}, these
resonances were also proposed to be $qq\bar{q}\bar{q}$
states~\cite{Jaffe}, hybrid $q\bar{q}$/meson--meson systems~\cite{Beveren} or even quark--less
gluonic hadrons~\cite{Jaffe1}.}
Among many theoretical 
investigations~\cite{kaminski,Baru,Teige,Bugg,Fu} the $K^+K^-$ interaction was studied
also experimentally in the $pp \to ppK^+K^-$ reaction with COSY--11 and ANKE detectors operating
at the COSY synchrotron in J\"{u}lich, Germany~\cite{wolke,quentmeier,winter,anke,Ye,f0,disto}. 
The experimental data collected systematically
below~\cite{wolke,quentmeier,winter,anke,Ye} and above~\cite{disto} the $\phi$ meson threshold reveal
a significant enhancement in the shape of the excitation function near the kinematical
threshold, which  may be due to the final state interaction (FSI) in the $ppK^+K^-$ system.
The indication of the influence of the $pK^-$ final state interaction was found in both \mbox{COSY--11}
and ANKE data in the ratios of the differential cross sections as a function of the $pK$ and
the $ppK$ invariant masses,
\begin{eqnarray}
\nonumber
R_{pK} &= \frac{\mathrm{d}\sigma/\mathrm{d}M_{pK^{-}}}{\mathrm{d}\sigma/\mathrm{d}M_{pK^{+}}}~,\\
\nonumber
R_{ppK} &= \frac{\mathrm{d}\sigma/\mathrm{d}M_{ppK^{-}}}{\mathrm{d}\sigma/\mathrm{d}M_{ppK^{+}}}~,
\end{eqnarray}
where a significant enhancement in the region of both the low $pK^-$ invariant mass $M_{pK^{-}}$
and the low $ppK^-$ invariant mass $M_{ppK^{-}}$ is
observed~\cite{anke,PhysRevC}.
\begin{figure}
\centering
\includegraphics[width=0.29\textwidth,angle=0]{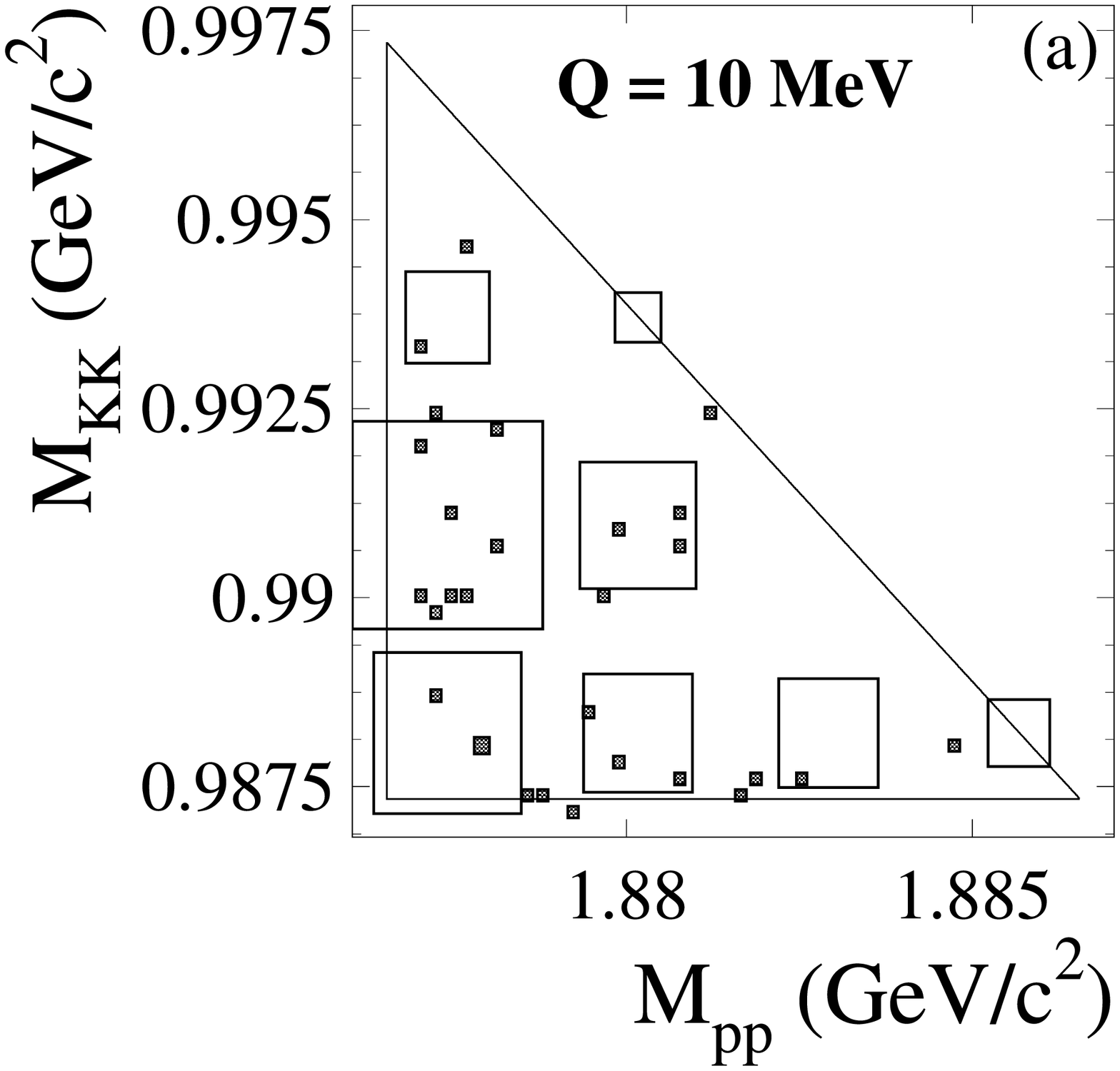}
\includegraphics[width=0.29\textwidth,angle=0]{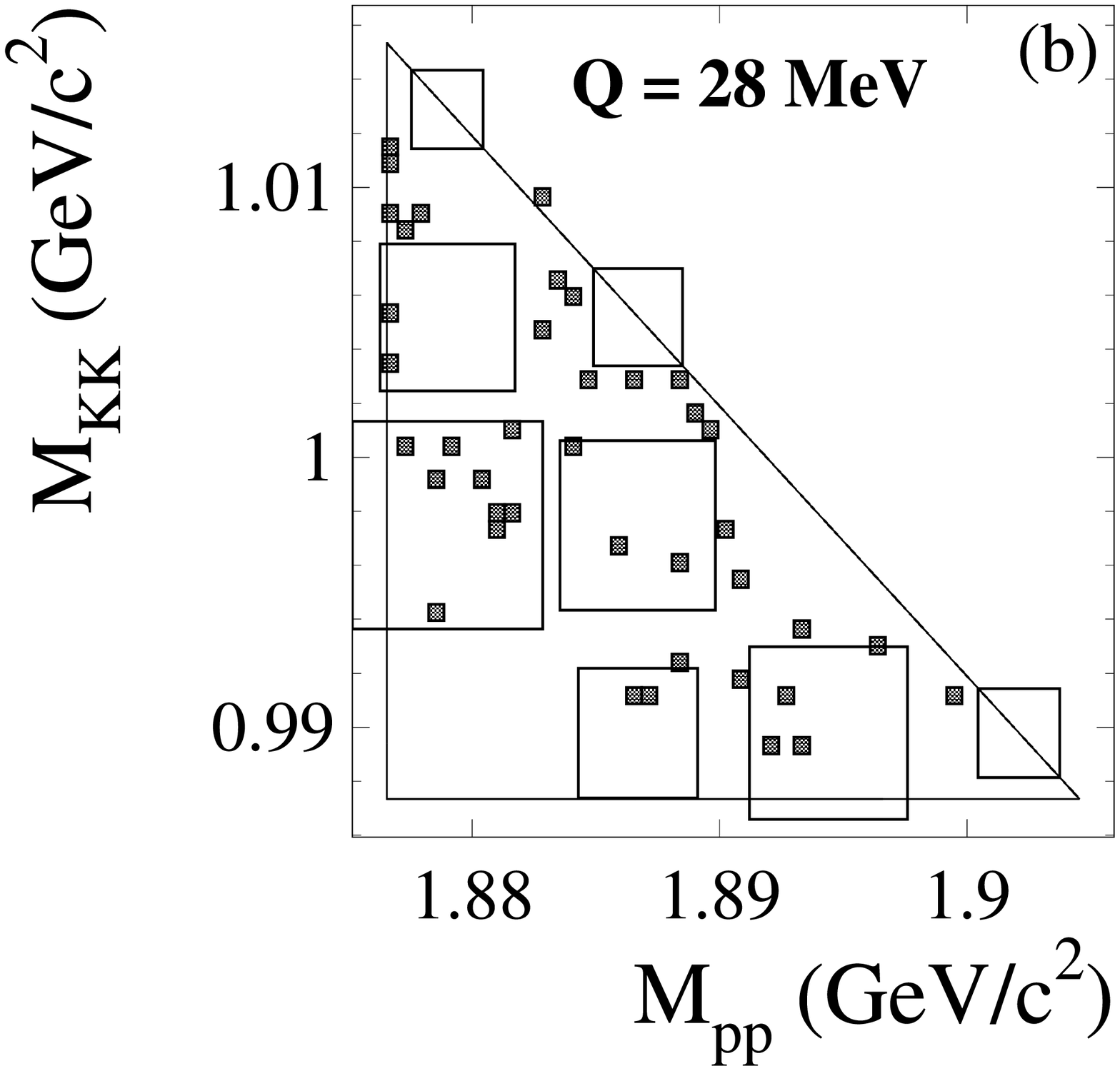}

\caption{
Experimental Goldhaber plots for the $pp\rightarrow ppK^{+}K^{-}$ reaction. 
The solid lines of the triangles show the 
kinematically allowed boundaries. 
Individual events are shown in (a) and (b) as black points. 
The superimposed squares represent the same distributions but binned into intervals of
$\Delta$M~=~2.5~MeV/c$^{2}$ ($\Delta$M~=~7~MeV/c$^{2}$) widths for an excess energy 
of Q~=~10 (28)~MeV, respectively.
The area of the square is proportional to the number of entries in a given interval.
The figure was adapted from~\cite{PhysRevC}.}
\label{fig:1}
\end{figure}
The phenomenological model based on the factorization of the final state interaction 
into interactions in the $pp$ and $pK^-$ subsystems, neglecting the $K^{+}K^{-}$ potential, 
does not  describe the whole experimental excitation function for the $pp \to ppK^{+}K^{-}$ reaction,
underestimating the data very close to the kinematical threshold~\cite{anke,wilkin}.
This indicates that in the low--energy region the influence of the $K^{+}K^{-}$
final state interaction may be significant~\cite{anke,PhysRevC,wilkin}. Motivated by this 
observation the COSY--11 Collaboration has recently estimated the scattering length of
the $K^{+}K^{-}$ interaction  based on the $pp \to ppK^{+}K^{-}$ reaction
measured at excess energies of Q~=~10 and 28~MeV~\cite{PhysRevC}. 
As a result of the analysis the $K^+K^-$ scattering length was determined based on the low--energy
proton--proton ($M_{pp}$) and $K^{+}K^{-}$ ($M_{KK}$)
invariant mass distributions (so--called Goldhaber plot) shown in Fig.~\ref{fig:1}~\cite{PhysRevC}.\\
In this article we combine the Goldhaber plot distribution established by the COSY--11 group
with the experimental excitation function~\cite{wolke,quentmeier,winter,anke,disto} near threshold 
and determine the $K^+K^-$ scattering length with better precision compared to the previous results.
We have also extracted the effective range of the $K^+K^-$ interaction.  
\section{Description of the final state interaction in the $ppK^+K^-$ system}
\label{sec2}
As in the previous analysis~\cite{PhysRevC} we use the factorization ansatz proposed by
the ANKE group with an additional term describing the interaction in the $K^+K^-$
system. We assume that the overall enhancement factor originating from final state
interaction can be factorized into enhancements in the proton--proton, the two $pK^-$
and the $K^+K^-$ subsystems\footnote{In this model we neglect the $pK^{+}$ 
interaction since it is repulsive and weak~\cite{anke}.}:
\begin{eqnarray}
\nonumber
F_{FSI} = F_{pp}(k_{1}) \times F_{p_{1}K^-}(k_{2}) \times F_{p_{2}K^-}(k_{3})\\
\times F_{K^+K^-}(k_{4})
\label{row1}
\end{eqnarray}
where $k_{j}$ stands for the relative momentum of particles in the corresponding
subsystem~\cite{PhysRevC}. The proton--proton scattering amplitude was taken
into account using the following parametrization:
\[F_{pp} =
  \frac{e^{i\delta_{pp}({^{1}\mbox{\scriptsize S}_{0}})} \cdot
        \sin{\delta_{pp}({^{1}\mbox{S}_0})}}
       {Ck_{1}}~,\]
where $C$ stands for the square root of the Coulomb pe\-ne\-tra\-tion factor~\cite{pp-FSI}.
The parameter $\delta_{pp}({^{1}\mbox{S}_0})$ denotes the phase shift 
calculated according to the modified Cini--Fubini--Stanghellini formula with
the Wong--Noyes Coulomb correction~\cite{noyes995,noyes465,naisse506}.
Factors describing the enhancement originating from the interaction in the $pK^-$ subsystems
are parametrized using the scattering length approximation:
\begin{eqnarray}
\nonumber
F_{pK^{-}}~=~\frac{1}{1~-~i~k~a_{pK^-}}~.
\label{F_pK}
\end{eqnarray}
The $pK^{-}$ scattering length $a_{pK^{-}}$ was estimated both theoretically
~\cite{Oller:2000fj,Ivanov:2003hn,Borasoy:2004kk,Oller:2005ig,Shevchenko:2006xy,Guo:2012vv},
and experimentally based mainly on the kaonic hydrogen atom measurements~\cite{Borasoy:2006sr,Martin}.
As shown in Ref.~\cite{Yan:2009mr} different approaches result in a slightly
different $a_{pK^{-}}$ values. Therefore, in our analysis we have assumed the
$pK^{-}$ scattering length to be equal to the mean of all the values from
elaborations
~\cite{Oller:2000fj,Ivanov:2003hn,Borasoy:2004kk,Oller:2005ig,Shevchenko:2006xy,Guo:2012vv,Borasoy:2006sr,Martin}
summarized in Ref.~\cite{Yan:2009mr}: $a_{pK^-} = (-0.65 + 0.78i$)~fm.\\
The $K^+K^-$--FSI was parametrized using the effective range expansion:
\begin{eqnarray}
\nonumber
F_{K^{+}K^{-}}=\frac{1}{\frac{1}{a_{K^+K^-}}+\frac{b_{K^+K^-}k^2_{4}}{2}-ik_{4}},
\label{F_KKb}
\end{eqnarray}
where $a_{K^+K^-}$ and $b_{K^+K^-}$ are the scattering length and the effective range
of the $K^{+}K^{-}$ interaction, respectively. We have performed a fit to
the experimental data treating $a_{K^+K^-}$ and $b_{K^+K^-}$ as free parameters.
Moreover, we have repeated the analysis for every quoted $a_{pK^-}$
to check, how their different values change the result. This allowed us also
to estimate the systematic error due to the $pK^{-}$ scattering length used in the estimation
of $a_{K^+K^-}$ and $b_{K^+K^-}$.\\
It is worth mentioning, that there is a similar phenomenological model of
the $K^{+}K^{-}$ final state interaction which takes into account the elastic
and charge--exchange interaction allowing for the $K^{0}K^{0}\rightleftharpoons K^{+}K^{-}$
transitions. This FSI should generate a significant cusp effect in the $K^{+}K^{-}$
invariant mass spectrum near the $K^{0}K^{0}$ threshold (details can be found in~\cite{dzyuba}).
Another contribution to this effect may be also generated by the kaons rescattering 
to scalars, eg.: $KK \to f_0(980) \to KK$ and $KK \to a_0(980) \to KK$.
\begin{figure*}
\centering
  \includegraphics[width=0.33\textwidth,angle=0]{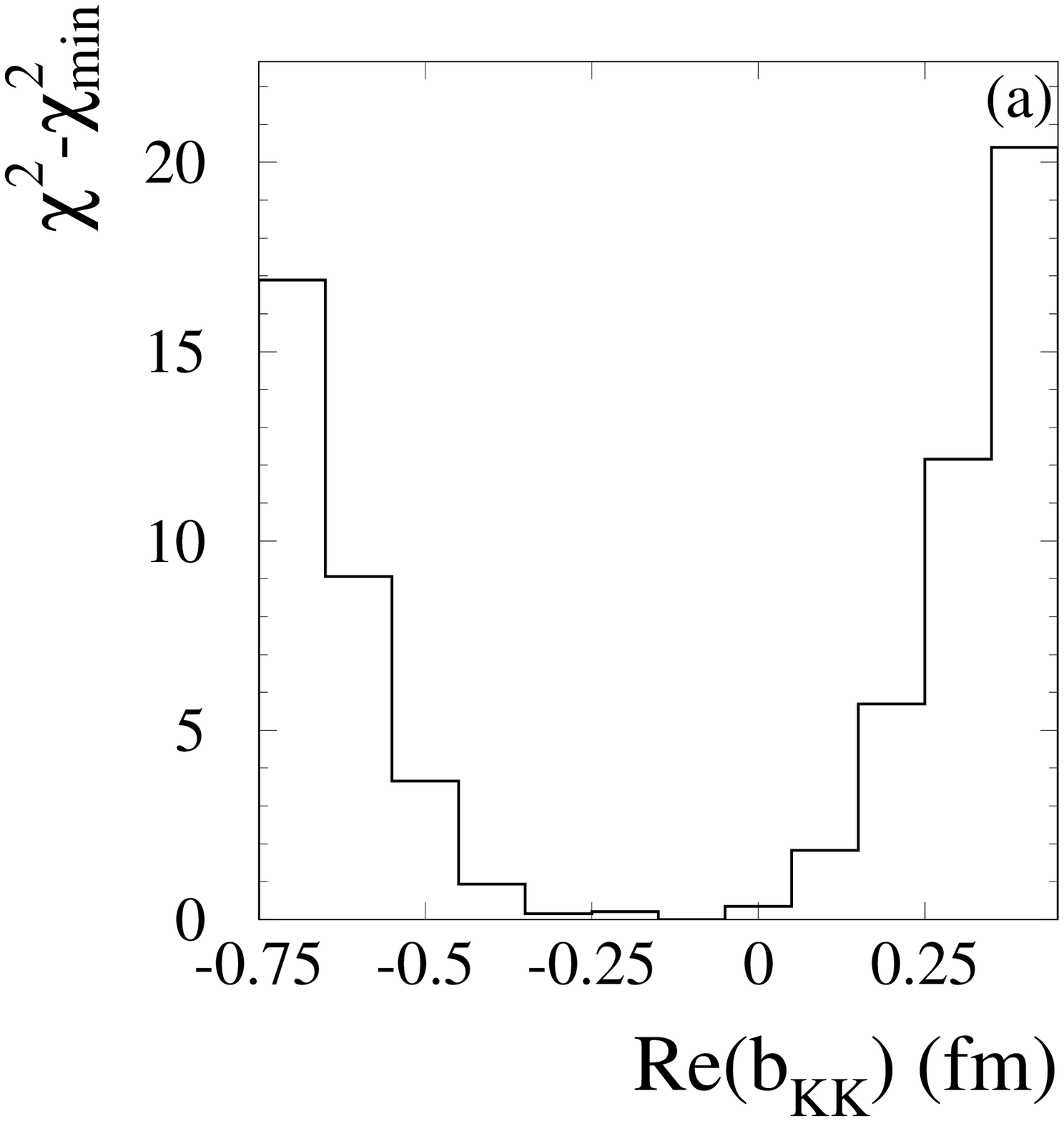}
 \includegraphics[width=0.33\textwidth,angle=0]{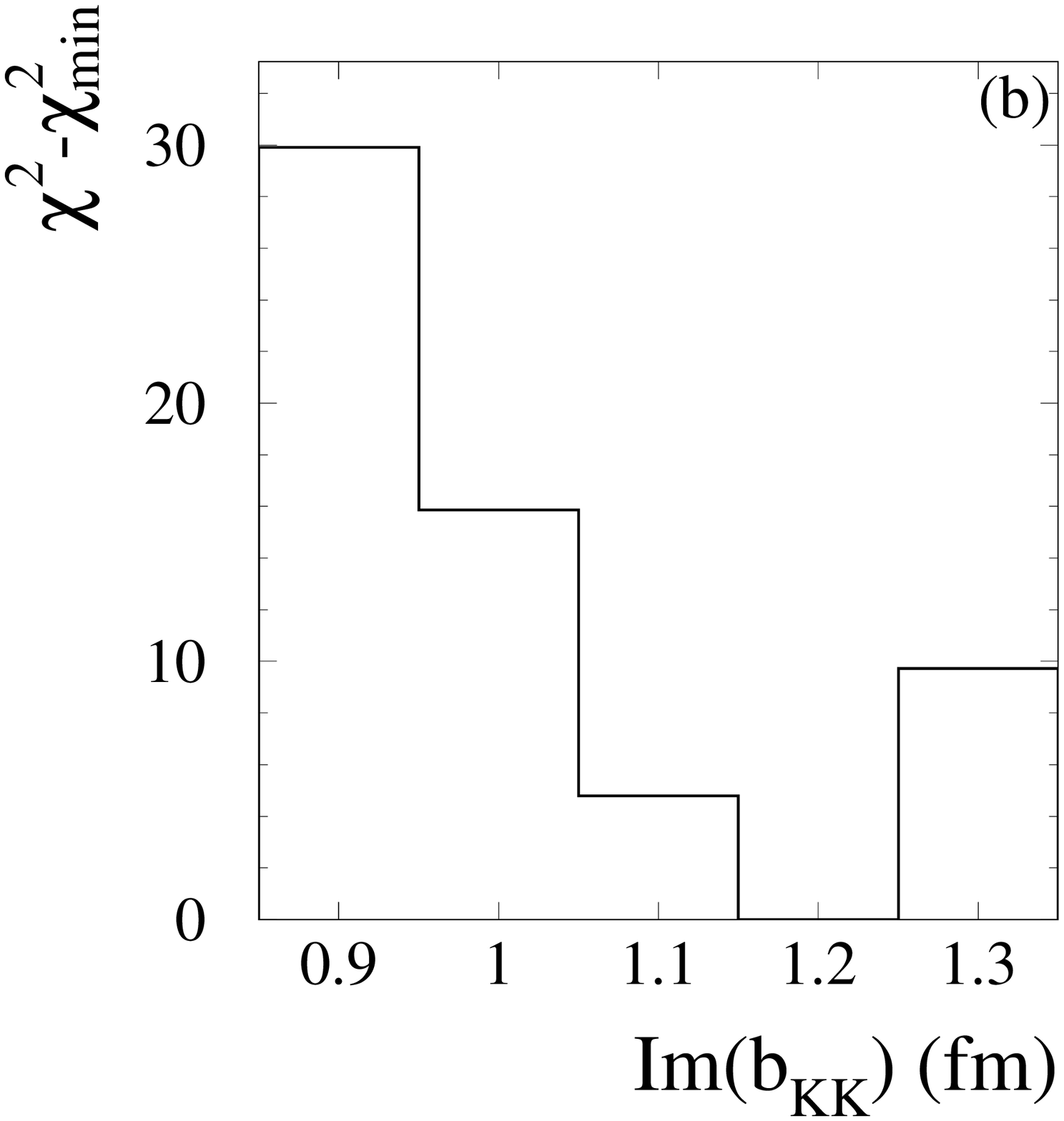}
  \includegraphics[width=0.33\textwidth,angle=0]{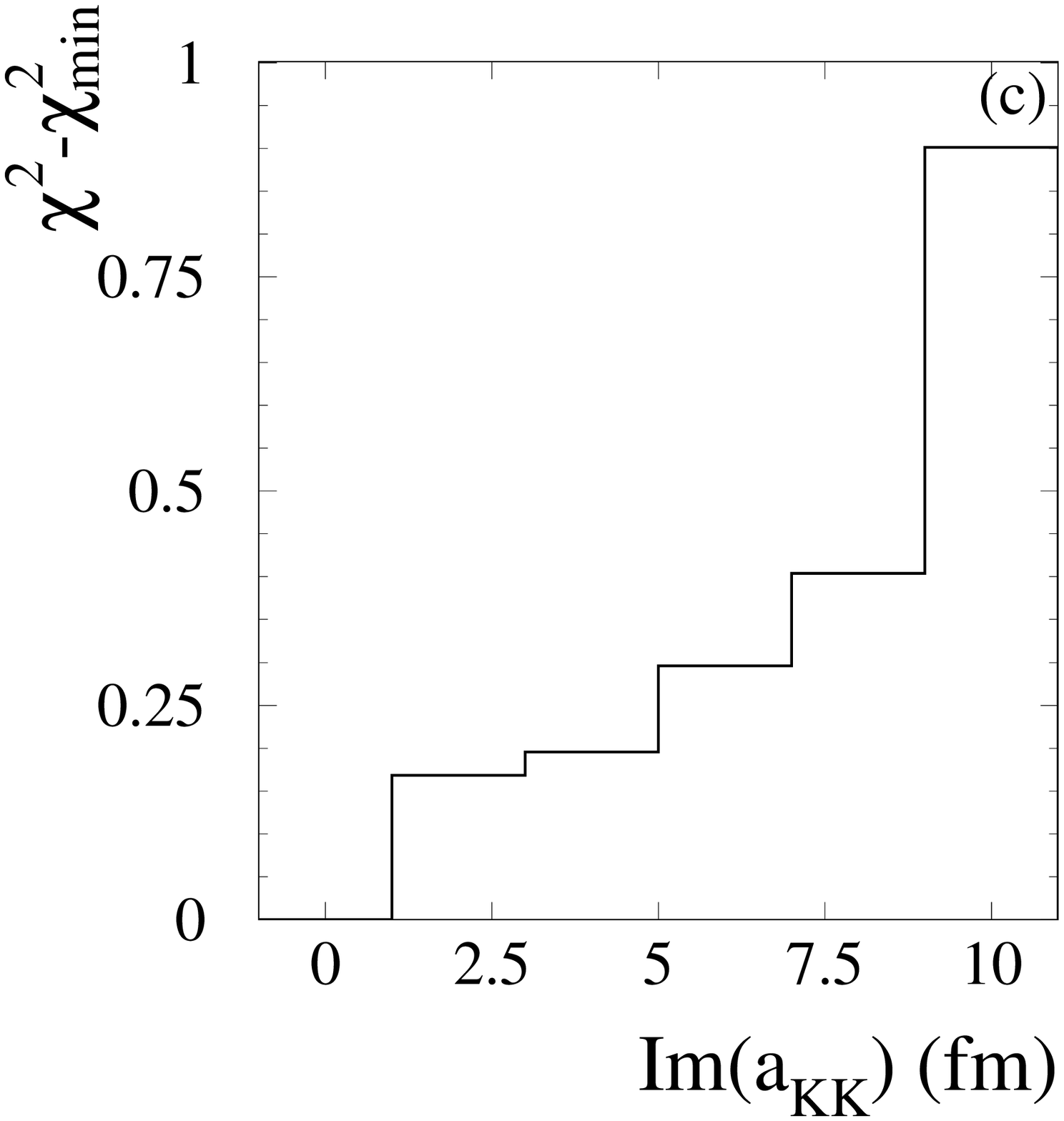}
  \includegraphics[width=0.33\textwidth,angle=0]{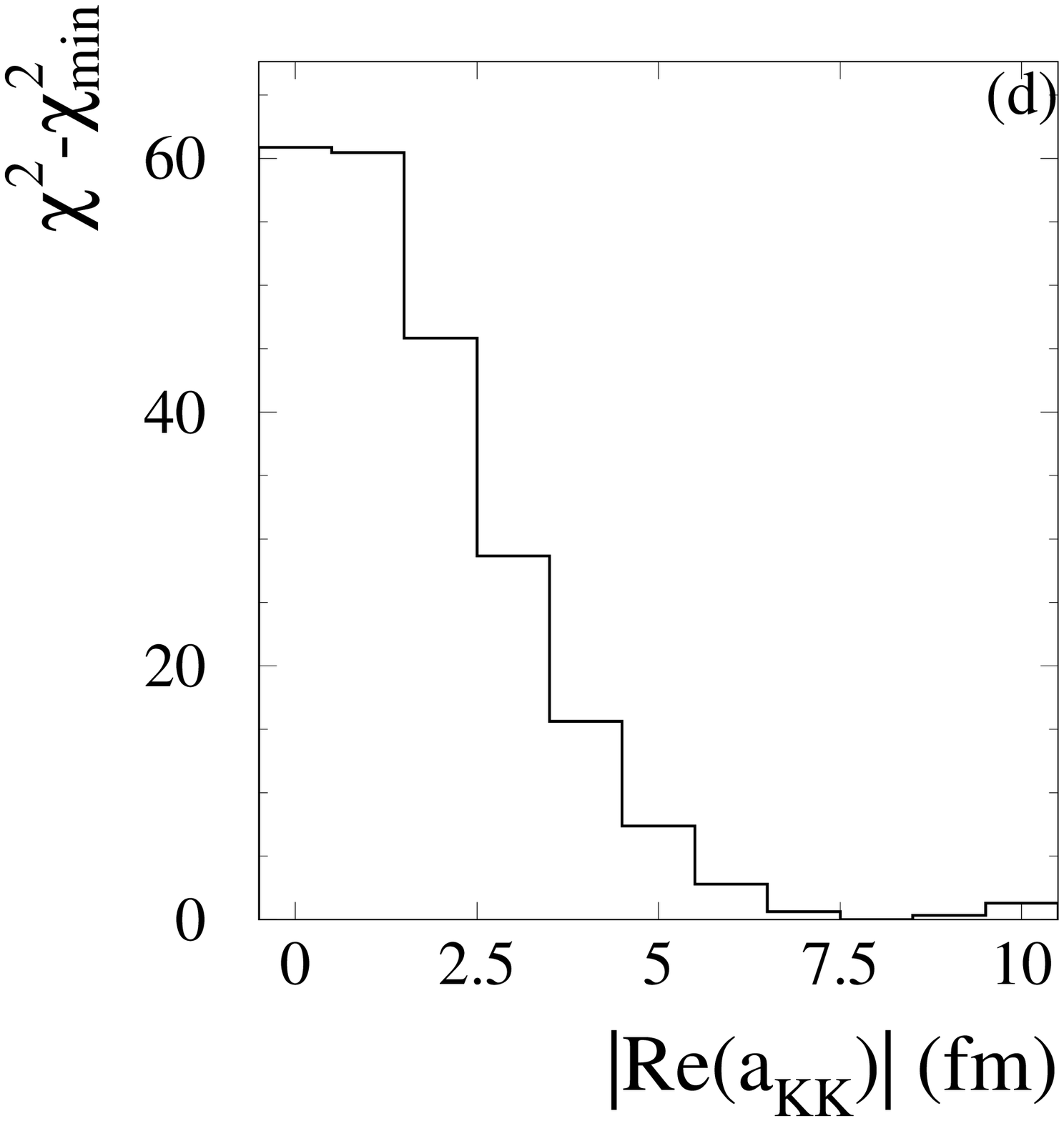}
\caption{$\chi^{2}$~-~$\chi^{2}_{min}$ distribution as a function of:
(a) $\mathrm{Re}(b_{K^{+}K^{-}})$, (b) $\mathrm{Im}(b_{K^{+}K^{-}})$,
(c) $\mathrm{Im}(a_{K^{+}K^{-}})$ and (d) $|\mathrm{Re}(a_{K^{+}K^{-}})|$.\\
$\chi^{2}_{min}$ denotes the absolute minimum with respect to parameters
$\alpha$, $\mathrm{Re}(b_{K^{+}K^{-}})$, $Im(b_{K^+K^-})$, $|\mathrm{Re}(a_{K^{+}K^{-}})|$,\\
and $Im(a_{K^+K^-})$.}
\label{fig:2}
\end{figure*}
However, the ANKE data can be described well without introducing the cusp
effect~\cite{dzyuba}, thus we neglect it in this analysis. We also cannot
distinguish between the isospin I~=~0 and I~=~1 states of the $K^+K^-$ system.
However, as pointed out in~\cite{dzyuba}, the production with I~=~0 is
dominant in the $pp\to ppK^+K^-$ reaction independent of the exact values
of the scattering lengths.\\
In the fit we do not take into account influence of the $f_0$(980) and $a_0$(980) production.
There exist only very rough experimental estimates of upper limits for production of these resonances in the $N-N$
collisions. In fact, up to now in these reactions there has not been found any signal of these particles.
The theoretical estimations result in negligible cross sections for the $pp \to f_0 pp \to K^+K^- pp$
resonant contribution with respect to the non-resonant one~\cite{Bratkovskaya} (the upper limit of the
cross section for this reaction is estimated to be about $1 \times 10^{-4}$~nb/MeV at $Q=5$~MeV
and $4 \times 10^{-2}$~nb/MeV at 50 MeV~\cite{Bratkovskaya}).
Also the branching ratios of $f_0$(980) and $a_0$(980) are very poorly known. However, according to the PDG
$a_0$(980) dominantly decays to $\eta\pi^0$ and the $\pi\pi$ channel is dominant for the $f_0$(980) meson~\cite{pdg2012}.
Thus, the $f_0$ resonance contribution to the near threshold $pp \to ppK^+K^-$ reaction is expected to be negligible.
Moreover, regarding the $a_0$(980) resonance, following Ref.~\cite{dzyuba} the $K^+K^-$ pairs are produced in
proton-proton collisions mainly with isospin I~=~0. Thus, $a_0$(980) would have to decay to $K^+K^-$
through isospin violation, which is an additional suppressing factor. 
According to Ref.~\cite{Bratkovskaya} for energies up to $Q=115$~MeV (DISTO measurement~\cite{disto})
the production of resonant $K^+K^-$ pairs should not produce any significant enhancement in the $K^+K^-$ invariant
mass.
\section{Determination of the $K^+K^-$ scattering length and effective range}
\label{sec3}
\begin{figure}
\centering
  \includegraphics[width=0.48\textwidth,angle=0]{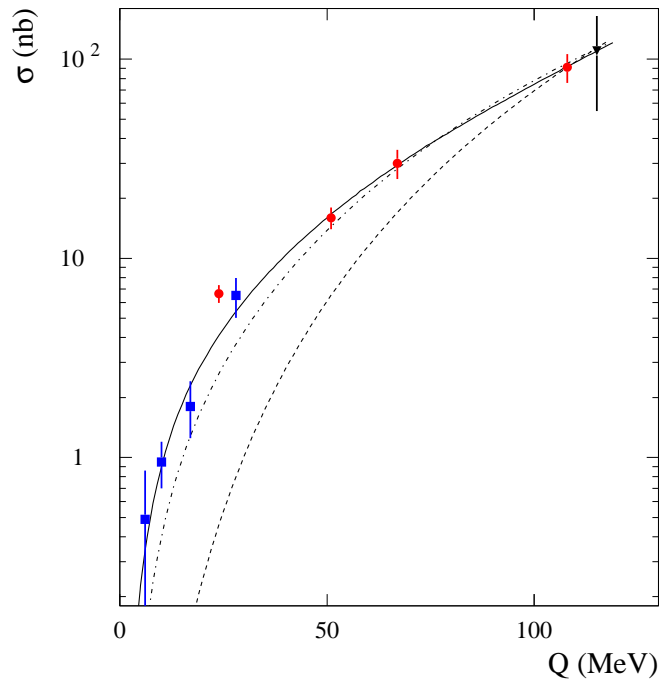}
\caption{(Color online) Excitation function for the $pp\rightarrow ppK^{+}K^{-}$ reaction.
Triangle and circles represent the DISTO and ANKE measurements, respectively~\cite{anke,anke_last,disto}.
The squares are results of the COSY--11~\cite{wolke,quentmeier,PhysRevC} measurements.
The dashed curve represents the energy dependence
obtained assuming that the phase space is homogeneously and isotropically populated,
and there is no interaction between particles in the final state.
Calculations taking into account the $pp$ and
$pK^-$ FSIs are presented as the dashed--dotted curve. 
The dashed and dashed--dotted curves are
normalized to the DISTO data point at Q~=~114~MeV. 
Solid curve corresponds to the result obtained taking into account
$pp$, $pK$, and $K^+K^-$ interactions parametrized with the effective range approximation.
These calculations were obtained using the scattering length $a_{K^{+}K^{-}}$ and effective
range $b_{K^+K^-}$ as obtained in this work. The latest data point measured by the ANKE group
was published recently~\cite{anke_last}, and thus it was not taken
into account in the fit.}
\label{fig:4}
\end{figure}

In order to estimate the strength of the $K^+K^-$ interaction
the experimental Goldhaber plots, determined at excess energies of Q~=~10 and 28~MeV, together with the total
cross sections measured near the threshold were compared to the results of the Monte Carlo
simulations treating the $K^+K^-$
scattering length $a_{K^{+}K^{-}}$ and effective range $b_{K^+K^-}$ as unknown complex
parameters.
To determine $a_{K^{+}K^{-}}$ and $b_{K^+K^-}$ we have constructed the following
$\chi^{2}$ statistics:
\begin{eqnarray}
\nonumber
\chi^2\left(a_{K^+K^-},b_{K^+K^-},\alpha\right) = \sum_{i=1}^{8}\frac{\left(\sigma_{i}^{expt}
- \alpha\sigma_{i}^{m}\right)^2} {\left(\Delta\sigma_{i}^{expt}\right)^2}\\
 +2\sum_{j=1}^{2}\sum_{k=1}^{10} \, [\beta_{j} N_{jk}^s - N_{jk}^e +  N_{jk}^e \,
{\mathrm{ln}}(\frac{N_{jk}^e}{\beta_{j} N_{jk}^s})] ,
\label{eqchi2_mh}
\end{eqnarray}
where the first term was defined following the Neyman's $\chi^{2}$ statistics,
and accounts for the excitation
function near the threshold for the $pp \to ppK^{+}K^{-}$ reaction.
$\sigma_{i}^{expt}$ denotes the $i$th experimental total cross section
measured with uncertainty $\Delta\sigma_{i}^{expt}$ and $\sigma_{i}^{m}$
stands for the calculated total cross section normalized with a factor $\alpha$
which is treated as an additional parameter of the fit. 
$\sigma_{i}^{m}$ was calculated for each excess energy $Q$ as a phase space integral
over five independent invariant masses~\cite{nyborg}:
\begin{eqnarray}
\nonumber
\sigma^{m}=\int\frac{\pi^{2}\left|M\right|^{2}}{8s\sqrt{-B}}~\mathrm{d}M^{2}_{pp}\mathrm{d}M^{2}_{K^{+}K^{-}}
\mathrm{d}M^{2}_{pK^{-}}\mathrm{d}M^{2}_{ppK^{-}}\mathrm{d}M^{2}_{ppK^{+}}.
\label{goldhaber3}
\end{eqnarray}
Here $s$ denotes the square of the total energy of the system determining the value
of the excess energy, and $B$ is a function of the invariant masses with the exact
form to be found in Nyborg's work~\cite{nyborg}. The amplitude for the process
$|M|^2$ contains the FSI enhancement factor defined in Eq.~(\ref{row1}) 
and it 	depends on the parameters $a_{K^{+}K^{-}}$ and $b_{K^{+}K^{-}}$.
The second term of Eq.~(\ref{eqchi2_mh}) corresponds to the Poisson likelihood
chi--square value~\cite{baker} describing the fit to the Goldhaber plots determined at excess energies
$Q = 10$~MeV ($j$~=~1) and $Q = 28$~MeV ($j$~=~2) using COSY--11 data~\cite{PhysRevC}.
$N_{jk}^e$ denotes the number of events in the $k$th 
bin of the $j$th experimental
Goldhaber plot, and $N_{jk}^s$ stands for the content of the same bin in the simulated
distributions. $\beta_{j}$ is a normalization factor which 
is fixed by values of the fit parameters $\alpha$ and $a_{K^{+}K^{-}}$.
It is defined for the $j^{th}$ excess energy
as the ratio of the total number of events expected from the calculated total cross section
$\sigma_{j}^{m}(a_{K^{+}K^{-}})$
and the total luminosity $L_{j}$~\cite{winter},
to the total number of simulated $pp \to ppK^+K^-$ events $N_{j}^{gen}$:
\begin{eqnarray}
\nonumber
\beta_{j} = \frac{L_{j}\alpha\sigma_{j}^{m}}{N_{j}^{gen}}.
\label{beta}
\end{eqnarray} 
The $\chi^2$ distributions (after subtraction of the minimum value) for $F_{K^+K^-}$
taken in the effective range expansion are presented as a function of the real and
imaginary parts of $a_{K^+K^-}$ and $b_{K^+K^-}$ in Fig.~\ref{fig:2}. 
The best fit to the experimental data corresponds to
\begin{eqnarray}
\nonumber
\mathrm{Re}(b_{K^{+}K^{-}}) = -0.1 \pm 0.4_{stat} \pm~0.3_{sys}~\mathrm{fm}\\
\nonumber
\mathrm{Im}(b_{K^{+}K^{-}}) = 1.2^{~+0.1_{stat}~+0.2_{sys}}_{~-0.2_{stat}~-0.0_{sys}}~\mathrm{fm},\\
\nonumber
\left|\mathrm{Re}(a_{K^{+}K^{-}})\right| = 8.0^{~+6.0_{stat}}_{~-4.0_{stat}}~\mathrm{fm}\\
\mathrm{Im}(a_{K^{+}K^{-}}) = 0.0^{~+20.0_{stat}}_{~-5.0_{stat}}~\mathrm{fm},
\label{chi2resultsB}
\end{eqnarray}
with a $\chi^2$ per degree of freedom of: $\chi^2/ndof = 1.30$. The statistical uncertainties
in this case were determined at the 70\% confidence level taking into account that we have varied five parameters
[$\alpha,\mathrm{Im}(a_{K^+K^-}),\mathrm{Re}(a_{K^+K^-}),\mathrm{Im}(b_{K^+K^-}),\mathrm{Re}(b_{K^+K^-})$].
Here uncertainties correspond to the range of values for which the $\chi^2$ of the fit
is equal to: $\chi^2 = \chi^2_{min} + 6.06$~\cite{james}. Systematic errors due to the uncertainties
in the assumed $pK^{-}$ scattering length were instead estimated as a maximal difference between
the obtained result and the $K^+K^{-}$ scattering length determined using different
$a_{pK^{-}}$ values\footnote{Due to the fact that in the case of
scattering length the systematic uncertainties are much smaller than the statistical ones we neglect
them in the final result.} quoted in Refs.~\cite{Guo:2012vv} and~\cite{Yan:2009mr}.
One can see, that the fit is in principle sensitive to both the scattering length and effective range,
however, with the available low statistics data the sensitivity to the scattering length is very weak.\\
Results of the analysis with inclusion of the interaction in the $K^+K^-$ system
described in this article are shown as the solid curve in Fig.~\ref{fig:4}.
One can see that the experimental data are described quite well over the whole energy range.
\section{Conclusions}
\label{secEnd}
We have performed a combined analysis of both total and differential
cross section distributions for the $pp\rightarrow ppK^{+}K^{-}$
reaction in view of the $K^+K^-$ final state interaction. In the analysis we have used
a factorization proposed by the ANKE group with an additional term describing
interaction in the $K^+K^-$ system, without distinguishing between the isospin
I~=~0 and I~=~1 states. We have also neglected a possible charge exchange
interaction leading to a cusp effect in the $K^{+}K^{-}$ invariant mass spectrum,
since taking it into account would require much more precise data~\cite{dzyuba}.
The $K^+K^-$ enhancement factor was parametrized using the effective range expansion.
Fit to experimental data is very weakly sensitive to $a_{K^+K^-}$, but allows us to
estimate for the first time the effective range of the $K^+K^-$--FSI.\\
All studies of the $pp \to ppK^+K^-$ reaction near threshold
~\cite{wolke,quentmeier,winter,anke,Ye} reveal that in the $ppK^+K^-$ system the interaction
between protons and the $K^-$ meson is dominant, and $a_{K^{+}K^{-}}$ is relatively small.
It seems that this reaction is driven by the $\Lambda(1405)$ production
$pp \to K^+\Lambda(1405) \to ppK^+K^-$ rather than by the scalar mesons~\cite{dzyuba}, which may,
however, contribute to the observed cusp effect by rescattering of kaons.
Thus, for precise determination of the kaon--antikaon scattering length we will need 
higher statistics, which can be available at, \textit{e.g.}, the ANKE experiment at
COSY~\cite{michael} or less
complicated final states like $K^+K^-\gamma$ or $K^0\overline{K^0}\gamma$, where only kaons
interact strongly. These final states can be studied for example via 
the $e^+e^- \to K^+K^-\gamma$ or $e^+e^- \to K^0\overline{K^0}\gamma$
reactions with the KLOE--2 detector operating at the DA$\Phi$NE
$\phi$--factory~\cite{AmelinoCamelia}.
\begin{acknowledgments}
We are grateful to M.~Hartmann, J.~Ritman, 
and A.~Wirzba for reading the earlier version of the manuscript and for their
comments and corrections.
We acknowledge the support by the FFE grants of the Research Center J{\"u}lich,  
by the Foundation for Polish Science and by the Polish National Science Center through the Grants No.
2011/03/N/ST2/02652 and 2011/03/B/ST2/01847.
\end{acknowledgments}

\end{document}